\documentclass[10pt,twoside,BCOR7mm,DIV15,headinclude,footexclude,
               cleardoubleempty,idxtotoc]
{scrartcl}

\usepackage{natbib}
\usepackage[font=small,labelfont=bf]{caption}
\usepackage[english]{babel}
\usepackage{graphicx}
\usepackage{hyperref}
\usepackage{scrpage2}
\usepackage{ifthen}
\usepackage{booktabs}
\usepackage{amsmath}
\usepackage{amssymb}
\usepackage{multicol}
\usepackage{float}
\usepackage{hyperref}

\hypersetup{breaklinks=true
,colorlinks=true,linkcolor=black,urlcolor=blue
,citecolor=black}

\addto\captionsenglish{%
}

\makeatletter
\renewcommand{\@biblabel}[1]{}
\renewcommand{\@cite}[2]{%
{#1\ifthenelse{\boolean{@tempswa}}{,#2}{}}}
\makeatother
\ofoot{\thepage}
\ifoot{}

\setcapindent{0em}
\setheadsepline{1pt}

\setkomafont{pagehead}{\normalfont\sffamily}
\setkomafont{pagenumber}{\normalfont\rmfamily}

\makeatletter
\newcommand{\listofcontributions}{\@starttoc{con}}

\newcommand{\l@contribution} {\@dottedtocline{1}{1.5em}{2.3em}}
\makeatother

\newenvironment{contribution}{
\setcounter{section}{0}
\setcounter{figure}{0}
\setcounter{table}{0}
}{
\newpage
\lehead{}
\rohead{}
}

\def\apj{ApJ}%
\def\aap{A\&A}%
\def\mnras{MNRAS}%
\begin{document}

\setlength{\baselineskip}{2.5ex}

\begin{contribution}

\lehead{C.\ Kehrig et al.}

\rohead{Metal-poor WR galaxies with IFS}

\begin{center}
{\LARGE \bf PopIII-star siblings in IZw18 and WRs in metal-poor galaxies unveiled from integral field spectroscopy}\\
\medskip

{\it\bf C. \ Kehrig$^1$, J.M.\ {V{\'{\i}}lchez}$^1$, E.\ P\'erez-Montero$^1$, J.\ Iglesias-P\'aramo$^{1,2}$, J.\ Brinchmann$^3$, P.A.\ Crowther$^4$, F.\ Durret$^5$, D.\ Kunth$^5$ \& et al.}\\

{\it $^1$Instituto de Astrof{\'{\i}}sica de Andaluc{\'{\i}}a (IAA-CSIC), Spain}\\
{\it $^2$ Estaci\'on Experimental de Zonas Aridas (CSIC), Spain }\\
{\it $^3$ Leiden University, The Netherlands }\\
{\it $^4$ University of Sheffield, United Kingdom}\\
{\it $^5$ Institut d'Astrophysique de Paris, France}

\begin{abstract}
 Here we highlight our recent results from the IFS study of Mrk178, {\it the closest metal-poor WR galaxy},
 and of IZw18, {\it the most metal-poor star-forming galaxy known in the
 local Universe}. The IFS data of Mrk178 show the importance of
 aperture effects on the search for WR features, and the extent to
 which physical variations in the ISM properties can be detected.  Our
 IFS data of IZw18 reveal its entire nebular
 HeII$\lambda$4686-emitting region, and indicate for the first time that peculiar, very hot (nearly)
 metal-free ionizing stars (called here {\it PopIII-star siblings}) might hold the
 key to the HeII-ionization in IZw18.
\end{abstract}
\end{center}
\begin{multicols}{2}

\section{Introduction}

Studying the WR content and radiative feedback from WRs in metal-poor
star-forming (SF) galaxies is crucial to test evolutionary models for
massive stars at low metallicity (Z), where the disagreement between observations
and such models is stronger \citep[e.g.,][]{L14}. In this context, we
have initiated a program to investigate nearby low-Z WR galaxies using
integral field spectroscopy (IFS). IFS has many advantages in a study
of this kind, in comparison with long-slit
spectroscopy \citep[e.g.,][]{K08,epm15,epm13}. By means of IFS one
can find WRs where they were not detected before. Also, IFS
is a powerful technique to probe and solve issues related with aperture effects,
and allows a more precise spatial correlation between massive stars and nebular
properties \citep[e.g.,][hereafter K13]{K08,K13}.  As a part of this
program, we have obtained new IFS data of the low-Z
galaxies Mrk178 and IZw18, published in K13 and \citet[][hereafter
K15]{K15}, respectively. Below we summarize the main results from
these two works.

\section{IZw18: PopIII-star siblings as the source of HeII-ionization}

We performed new IFS observations of IZw18 using the PMAS IFU
\citep{R05} on the 3.5m telescope at CAHA (fig.\ref{kehrig.izw18.fig3}; K15). IZw18 is a nearby SF galaxy,
well known for its extremely low \mbox{Z$\sim$(1/40) Z$_{\odot}$}
\citep[e.g.,][]{jvm98}, and it is considered an excellent local analog
of primeval systems. Our IFS data reveal for the first time the entire
nebular HeII$\lambda$4686-emitting region (fig.\ref{kehrig.izw18.fig4}) and corresponding total HeII-ionizing
photon flux [Q(HeII)$_{obs}$] in IZw18. Narrow HeII emission in SF galaxies has
been suggested to be mainly associated with photoionization from WRs, but WRs cannot
satisfactorily explain the HeII-ionization in all cases, particularly
at lowest metallicities where HeII emission is observed to be
stronger \citep[e.g.,][]{G00,K04,sb12}. Why is studying the formation of
HeII emission relevant ? HeII emission indicates the presence of high
energy photons (E $\geq$ 54 eV), and 
HeII-emitters are apparently more frequent among high-redshift (z)
galaxies than for local objects \citep[e.g.,][]{K11,cassata13}. Narrow
HeII emission has been suggested as a good tracer of PopIII-stars (the
first very hot metal-free stars) in high-z galaxies
\citep[e.g.,][]{s03,p2015}; these stars are believed to have
contributed significantly to the reionization of the Universe, a
challenging subject in contemporary cosmology. However, the origin of narrow HeII
lines remains difficult to understand in
many nearby SF galaxies/regions \citep[e.g.,][]{K11,sb12}. So before
interpreting high-z HeII-emitters, it is crucial first to
understand the formation of HeII emission at low redshift. IZw18, as the most metal-poor HeII-emitter in the local
Universe, is an ideal object to perform this study.

Our observations combined with stellar model predictions point out
that conventional excitation sources (e.g.  single WRs, shocks, X-ray
binaries) cannot convincingly explain the total Q(HeII)$_{obs}$
derived for IZw18 \citep[e.g.,][]{mm05,crowther06,L14}. Other
mechanisms are probably also at work. If the HeII-ionization in IZw18
is due to stellar sources, these might be peculiar very hot
stars. Based on models of very massive O stars \citep[][]{KU02},
$\sim$ 10-20 stars with 300 M$_{\odot}$ at Z$_{IZw18}$ [or lower, down
to Z$\sim$(1/100) Z$_{\odot}$] can reproduce our total
Q(HeII)$_{obs}$ \citep[see also][]{dori}. However, the super-massive
star scenario requires a cluster mass much higher than the mass of the IZw18
NW knot (where the HeII region is located), and it would not be hard enough to explain the highest
HeII/H$\beta$ values observed. Also, stellar sources with 300
M$_{\odot}$ are not observed in IZw18 to date.  Next, as an
approximation of (nearly) metal-free stars in IZw18 -- the so-called
{\it PopIII-star siblings} -- we compared our observations with models
for rotating Z=0 stars \citep[][]{yoon12}, which 
reproduce our data better: $\sim$8-10 of such stars with M$_{ini}$=150
M$_{\odot}$ can explain the total Q(HeII)$_{obs}$ and the highest
HeII/H$\beta$ values observed.  The PopIII-star sibling scenario,
invoked for the first time in IZw18 by K15, goes in line with the
reported metal-free gas pockets in the HI envelope near the IZw18 NW
knot \citep[][]{L13}. These gas pockets could provide the raw material
for making such {\it PopIII-star siblings}.


\begin{figure}[H]
\begin{center}
\includegraphics[width=5.5cm,bb=3 56 1008 903,clip]{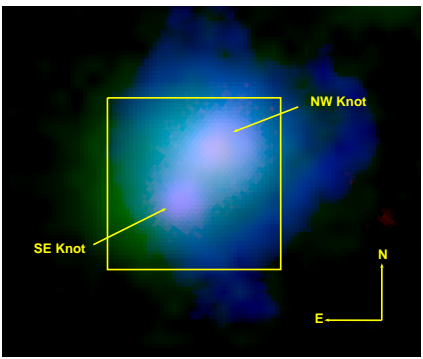} 
\caption{Color-composite image of IZw18. The box represents the FOV (16''$\times$16'') of the PMAS IFU over the galaxy main body which hosts the NW and SE knots (Figure taken from K15).
\label{kehrig.izw18.fig3}}
\end{center}
\end{figure}


\begin{figure*}[!t]
\begin{center}
\includegraphics[width=6cm,clip]{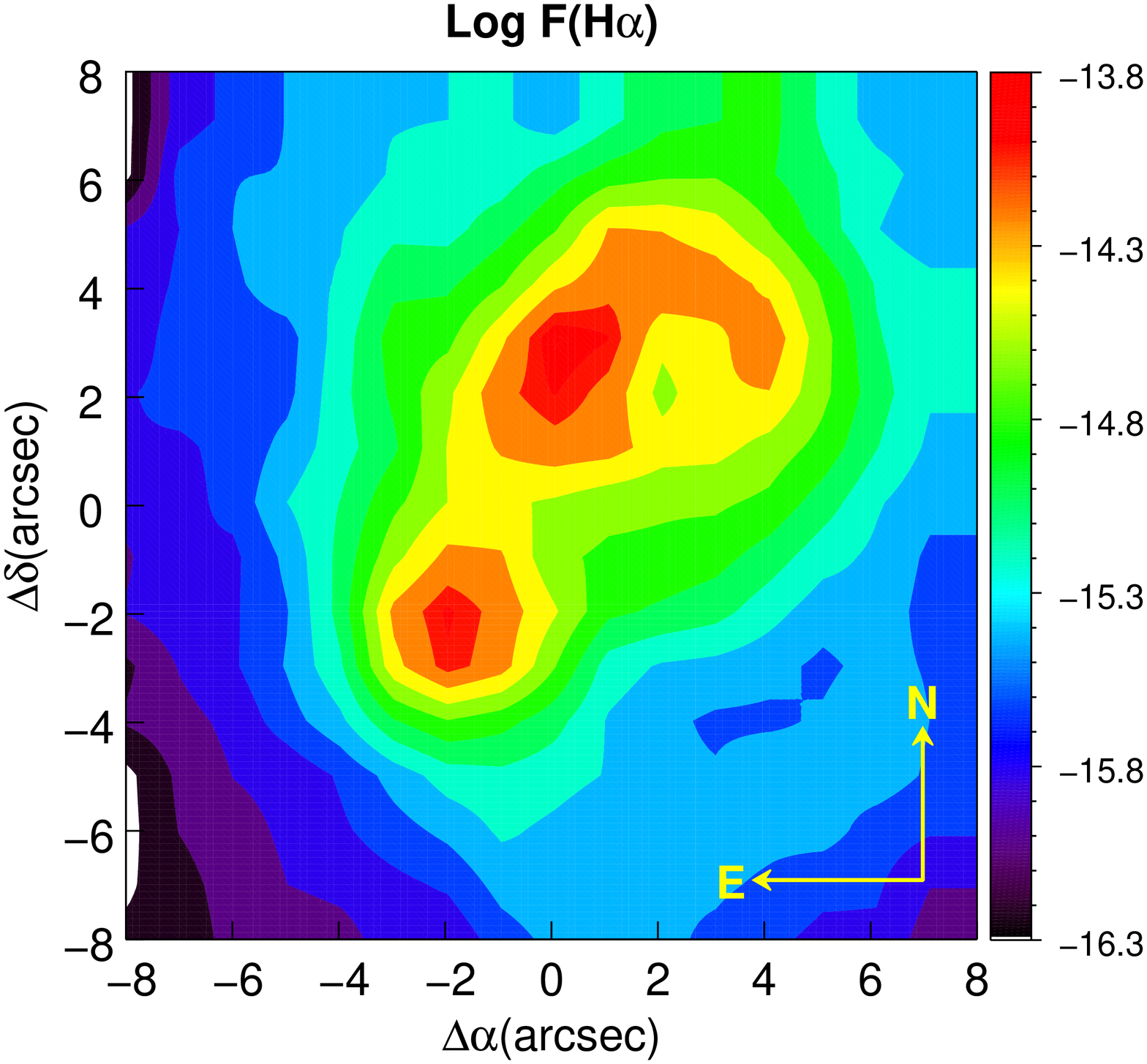}
\includegraphics[width=6cm,clip]{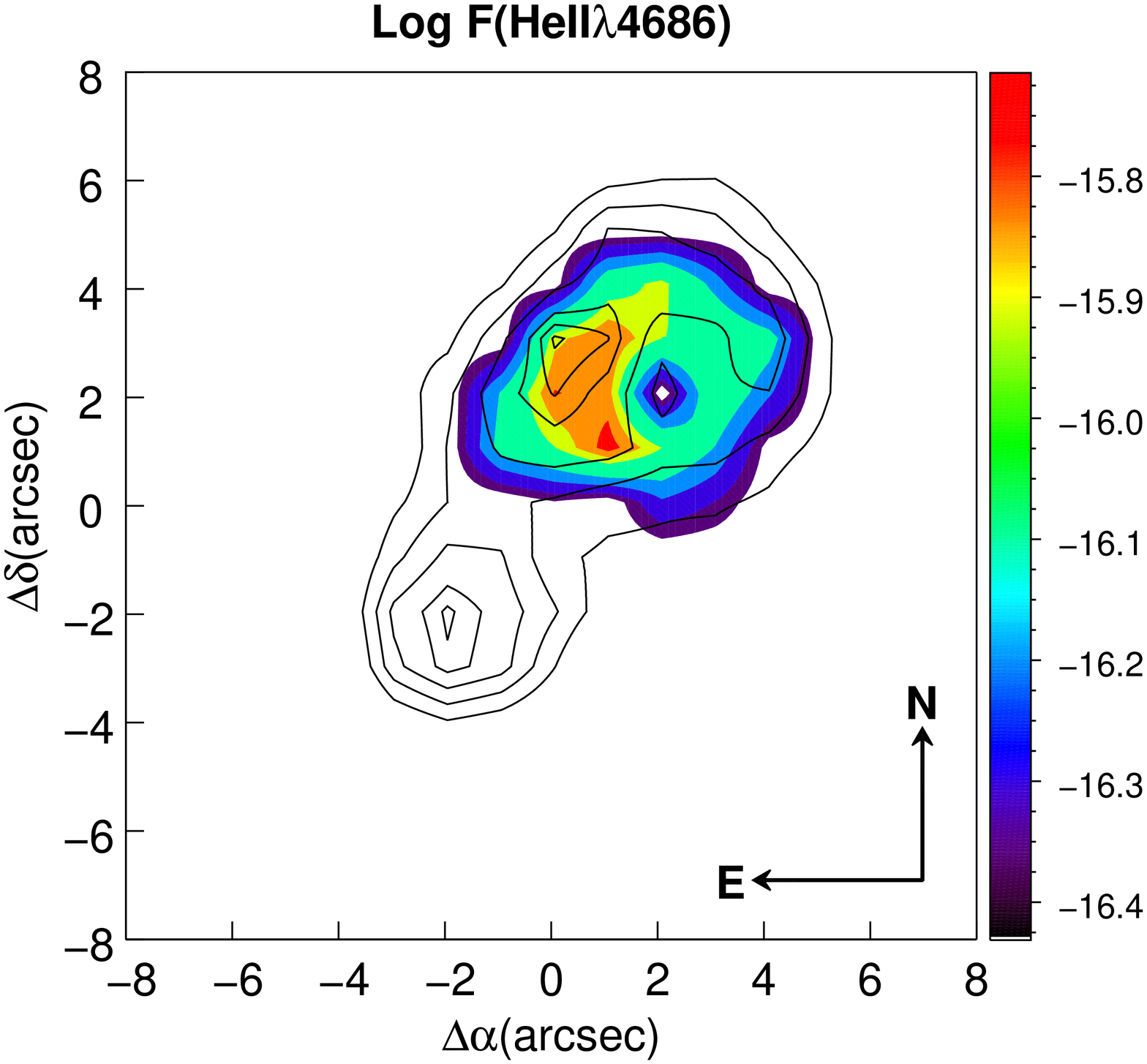} 
\caption{IZw18: intensity maps of H$\alpha$ ({\it left panel}) and nebular HeII$\lambda$4686 ({\it right panel}) in logarithmic scale; fluxes are in units of erg s$^{-1}$ cm$^{-2}$. The maps are presented as color-filled contour plots (Figure taken from K15).
\label{kehrig.izw18.fig4}}
\end{center}
\end{figure*}

\section{Mrk178}

In K13 we present the first optical IFS study of Mrk178, the
closest metal-poor WR HII galaxy. IFS data of Mrk 178 were obtained
with the INTEGRAL IFU at the 4.2m WHT. In this work we examine
the spatial correlation between its WRs and the neighbouring ionized
ISM. The strength of the broad WR features and its low metallicity
($\sim$ 1/10 Z$_{\odot}$) make Mrk178 an intriguing object. We have
detected the blue and red WR bumps in different locations across the
FOV ($\sim$ 300 pc$\times$230 pc) in Mrk178
(fig.\ref{kehrig.mrk178.fig1}).  The study of the WR content has been
extended, for the first time, beyond its brightest SF knot
(knot B in fig.\ref{kehrig.mrk178.fig1}) uncovering new WR star
clusters (knots A and C in fig.\ref{kehrig.mrk178.fig1}). Using
SMC/LMC template WRs \citep[][]{crowther06}, we empirically estimate a minimum of $\sim$
20 WRs in our Mrk178 FOV, which is already higher than that currently
found in the literature. Regarding the ISM abundances, localized N and
He enrichment, spatially correlated with WRs from knot B, is suggested by our
analysis.  Nebular HeII$\lambda$4686 emission is shown to be spatially
extended reaching well beyond the location of the WRs
(fig.\ref{kehrig.mrk178.fig2}). Shock ionization and X-ray
binaries are unlikely to be significant ionizing mechanisms since
Mrk178 is not detected in X-rays. The main excitation source of
HeII in Mrk178 is still unknown.


\begin{figure*}[!t]
\begin{center}
\includegraphics[width=6cm,clip]{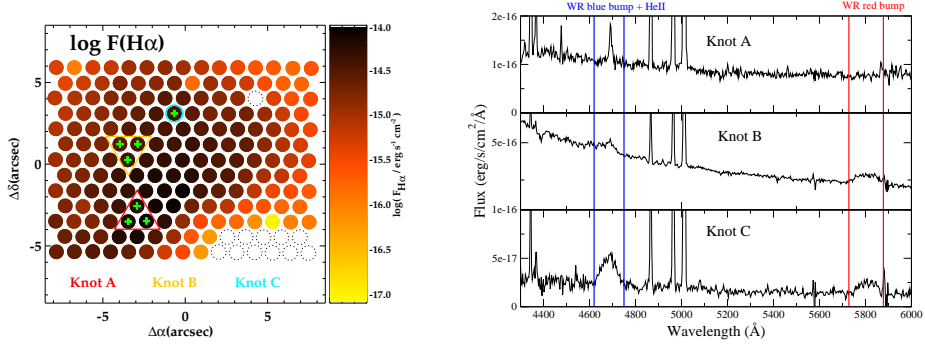}
\includegraphics[width=6cm,clip]{wrbumps.eps}
\caption{Mrk178: intensity map of H$\alpha$ emission line ({\it left panel}) and integrated spectrum for the 3 knots in which WR features are detected ({\it right panel}). The spectral range for both blue and red WR bumps are marked (K13). 
\label{kehrig.mrk178.fig1}}
\end{center}
\end{figure*}


\begin{figure*}[!t]
\begin{center}
\includegraphics[width=6cm,clip]{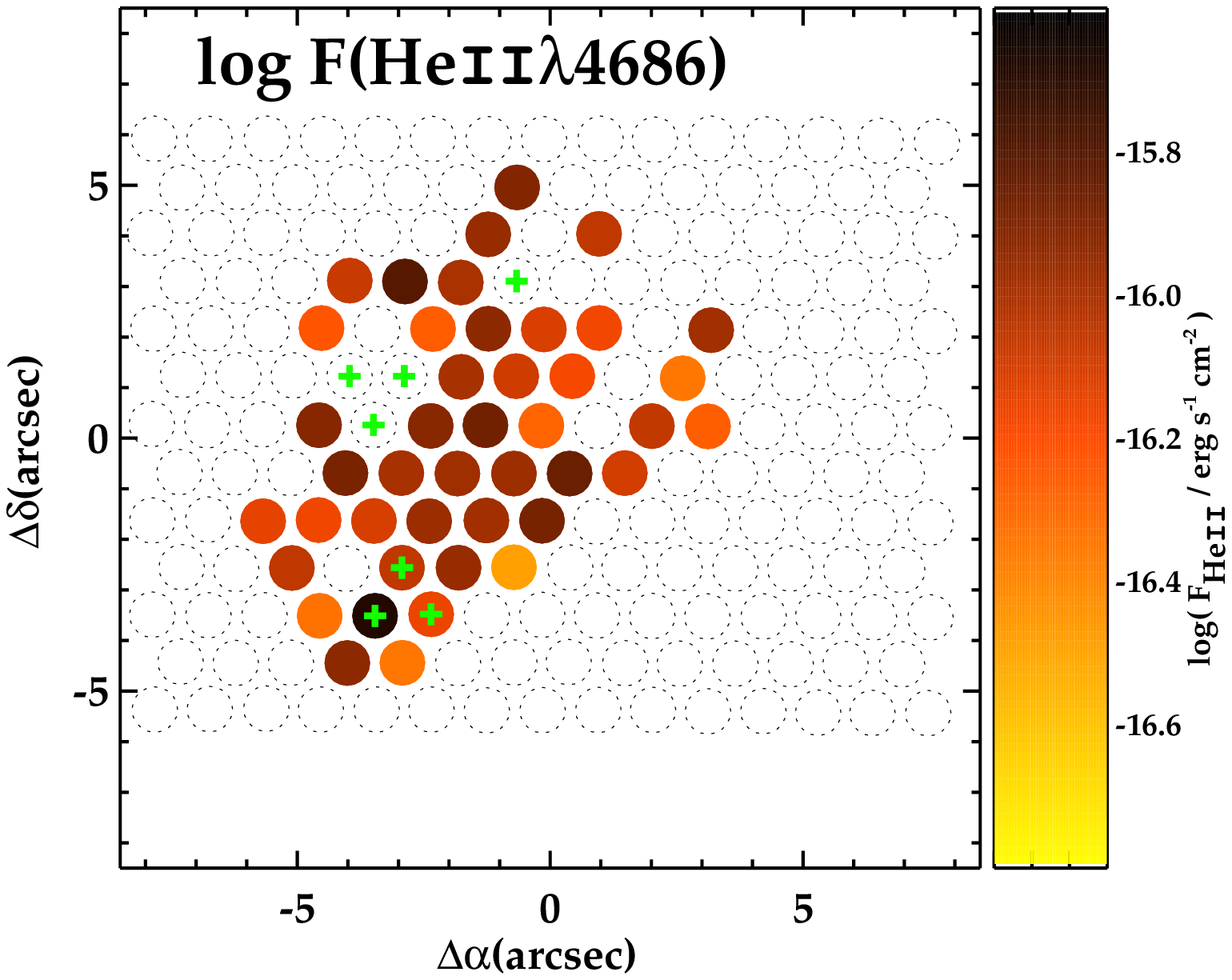}
\includegraphics[width=6cm,clip]{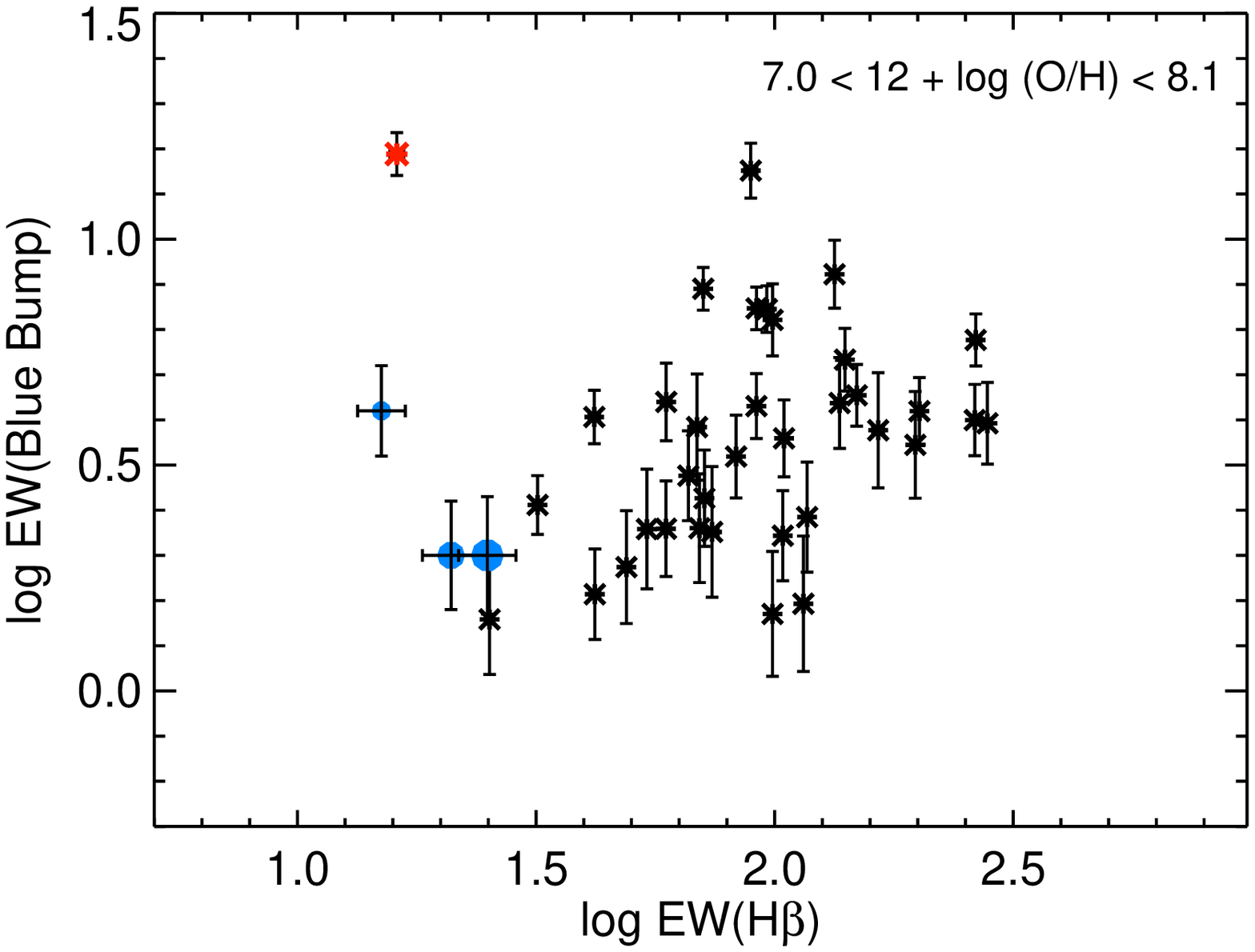} 
\caption{{\it Left panel}: map of nebular HeII$\lambda$4686 line; spaxels where
we detect WR features are marked with green crosses.
{\it Right panel}: EW(WR blue bump) vs EW(H$\beta$). Asterisks show values from SDSS DR7 for metal-poor WR galaxies; the red
one represents Mrk178. The three blue circles, from the smallest to the biggest one, represent the
5'', 7'' and 10'' diameter apertures from our IFU data centered at the SDSS fiber of Mrk178 (K13).
\label{kehrig.mrk178.fig2}}
\end{center}
\end{figure*}

From SDSS spectra of metal-poor WR galaxies, we found a 
too high EW(WR bump)/EW(H$\beta$) value for Mrk178, which is the
most deviant point in the sample (fig.\ref{kehrig.mrk178.fig2}).  Using our IFU
data, we showed that this curious behaviour is caused by
aperture effects, which actually affect, to some degree, the EW(WR
bump) measurements for all galaxies in Fig.\ref{kehrig.mrk178.fig2}. Also, we demonstrated that
using too large an aperture, the chance of detecting WR features
decreases, and that WR signatures can escape detection depending on
the distance of the object and on the aperture size. Thus, WR
galaxy samples constructed on a single fiber/long-slit
spectrum basis may be affected by systematic bias.


\end{multicols}

\end{contribution}


\end{document}